\documentclass[preprintnumbers,amsmath,amssymb]{revtex4}
\usepackage{amsmath} % need for subequations
\usepackage{graphicx} % for figures
\usepackage{color}

\pdfoutput=1
\begin{document}
\title{On the quantum ``self-teleportation'' probability of a human body}
\author{Massimiliano Sassoli de Bianchi}
\affiliation{Center Leo Apostel for Interdisciplinary Studies, Brussels Free University, Brussels, Belgium}
\email{msassoli@vub.ac.be}

\date{\today}

\begin{abstract}
\noindent The probability of quantum relocation of a human body, at a given distance, is estimated using two different methods, giving comparable results. Not only the obtained values for the probabilities are inconceivably small, but assumptions of a sci-fi nature are also necessary to ensure that they are not identical to zero. The notions of `non-spatiality' and `superselection rule' are also briefly discussed.
\end{abstract}
\maketitle

\section{Introduction}
\noindent Recently a science fiction writer asked me the following question: 
\\

\noindent\emph{What is the probability for an individual to suddenly vanish from one place and, one second after, reappear in another predetermined place, tens of kilometers away, according to the laws of quantum physics}? 
\\

\noindent He  also told me that a famous physicist (he had forgotten the name) used to pose such question to his students, so that it necessarily had to be a simple textbook problem. His interest in this question was that the protagonist of his story had to take advantage of this probability, no matter how infinitesimal (he was equipped with a futurist amplifier of probabilities), to ``transfer'' all of a sudden his body to a considerable spatial distance. 

Inspired by his curious quiz, which only apparently is a textbook one, I will try in this article to offer a few elements of clarification about some important concepts of quantum physics, in particular the concept of \emph{non-spatiality}, which I will illustrate by means of a simple metaphor. I will also provide two different estimates of the teleportation probability in question, on the basis of a number of simplifying assumptions, some of which will necessarily be of a sci-fi nature. Despite these assumptions, the values I will obtain are so small that they are almost impossible to conceive.

Not to create any misunderstanding, let me assert very clearly, from the beginning, that the probability that in \emph{normal conditions} an individual would disappear from one place and be teleported to another place is, according to the today known laws of quantum physics (and the author's personal understanding of them), exactly \emph{equal to zero}! In fact, as I will explain, and until evidence to the contrary, ordinary macroscopic bodies (such as our human bodies), in standard environmental conditions (for instance of temperature and pressure), do not obey the quantum laws. But before proceeding in my discussion, I have to face a little problem of terminology. 

The term ``teleportation'' is used in quantum physics to denote a very specific class of phenomena that have nothing to do with the nature of the question addressed to me by the science fiction writer~\cite{Bennet1993}. These phenomena describe the possibility of carrying information from one place to another, in ways that allow the construction of an exact duplicate of a given physical entity. This construction can be obtained only on the condition that the system of origin (i.e., the system that is to be duplicated) is altered, if not destroyed, in the process, since a well-known theorem, called the \emph{no-cloning theorem}, forbids to create a perfect clone of a quantum entity~\cite{Wootters1982, Dieks1982}.

Apart from this difficulty, the quantum teleportation, understood in the usual sense mentioned above, requires the preparation of special pairs of non-separated (entangled) systems that have to connect the two spatial regions between which the teleportation is to be produced, which for this reason is also called \emph{entanglement-assisted teleportation}. In other words, this form of quantum teleportation requires the presence of technological apparatuses tailored to the specificities of the entities to be teleported, and the execution of a series of operations that wil produce their destruction and reconstitution in the place of destination. This is not a spontaneous process, associated with probabilities, but a determinative process, which requires a specific technology to be implemented. 

Let me add that in the quantum teleportation only the information about the entity is transported, so as to allow its reconstruction, while nothing material is actually moved (except the carriers of information along an ordinary communication channel); and even though experiments of entanglement-assisted teleportation have already been successfully carried out (the current record is a teleportation over a distance of 143 km, between the Canary Islands of La Palma and Tenerife~\cite{Ma2012}), these remain so far limited to individual microscopic entities and finite-dimensional physical observables, such as the polarization of a photon, or the spin of an electron.

Having said that, and in order to avoid misunderstandings, I will use in the following the term ``quantum self-teleportation,'' or more simply \emph{self-teleportation}, to designate a hypothetical process of spatial relocation of a physical entity, to distinguish it from the aforementioned quantum teleportation assisted by entanglement. As previously emphasized, self-teleportation does not seem to be possible for macroscopic bodies, in standard conditions, as they only obey the laws of classical physics. Therefore, some additional sci-fi-like assumptions will be needed to explore this possibility and provide an estimate of the probability of such event, for a macroscopic entity like the body of a human being.

\section{Non-spatiality}

I will start by explaining a little better why a macroscopic body cannot behave like a microscopic entity. It is important to observe that macroscopic bodies, such as human bodies, or whatever ordinary objects, like rocks, grains of sand, etc., are \emph{spatial entities}. This means that they evolve while remaining within the so-called 3-dimensional Euclidean space. To clarify what I mean by this, I will use a simple metaphor. 

Imagine a swimmer in a pool. The pool's water corresponds to the 3-dimensional physical space, and the swimmer in it represents a macroscopic entity. If she wants to move from one point to another of the pool, that is, from one point in space to another point in space, she can only do so by swimming, and of course, due to the viscosity of water, the speed of her movement will be limited: she will not be able to exceed a determined maximum speed, which we can assume to be, say, of $2$~m/s. Thus, if we assume that the swimmer is located near the trampoline, and she wants to reach a point located at the center of the pool, say $10$~m away, this will take her $5$~s, if she can travel at the maximum possible speed.

Imagine now a child on the trampoline. In this metaphor the child is a microscopic entity, located outside of the pool's water, that is, outside the ordinary 3-dimensional physical space. Indeed, microscopic entities, when not organized into macroscopic aggregates, or when not interacting with macroscopic entities, are typically non-spatial entities~\cite{Aerts1998,Aerts1999,Sassoli2011,Sassoli2012,Sassoli2013,AertsSassoli2014,Sassoli2015}, not belonging to the water of the pool. The child, as a non-spatial entity, ``moves'' through another ``space,'' which in a sense is adherent to our physical space, and which in our metaphor is represented by the layer of air above the pool; and since the viscosity of  air is lower than that of  water, he will be able to do so with greater effective speed than the swimmer.

Suppose that the limit speed in the air, for the child, is of $10$~m/s, and that he is actually running at that speed on the trampoline, while in the process of diving. He will then be able to pass from the region of the trampoline to the region of the center of the pool in about $1$~s, which is something the swimmer is obviously unable to do. The interesting thing is that from the perspective of the swimmer, it is as if the diving child would appear out of nowhere in the middle of the pool, because he was not moving through the water, as the swimmer is forced to do, but through the air, which corresponds to a different layer of reality, of a non-ordinary kind, that we cannot directly perceive using our ordinary perceptual tools. 

I hope it is clear to everyone that a swimmer who is immersed in the water of the pool will never be able to move from one point to another as a diver (who is outside of it) can do. Similarly, a macroscopic body (the swimmer in our metaphor), being forced to move while remaining in the 3-dimensional physical space, will never be able to mimic the behavior of a microscopic entity, which instead is almost always outside of it (unless of course we would find a way to bring it out from that water, which is the sci-fi hypothesis we will have later on to consider). 

There are several ways to infer the mysterious non-spatiality of the microscopic entities. The simplest is to take seriously the uncertainty principle of Heisenberg. In fact, according to it, it is not possible to simultaneously determine  both the position and momentum of a microscopic entity. Therefore, it is not possible to solve the equations of motion (which require as an input both quantities), and as a consequence it is not possible to determine the spatial trajectory of the entity in question. This impossibility does not arise from the fact that we would lack some crucial information about the state of the entity (as the \emph{no-go theorems} about hidden-variable theories illustrate~\cite{Neumann1932,Bell1966,Gleason1957,Jauch1963,Kochen1967,Gudder1970}), which if we would possess would allow us to determine its trajectory: it is an impossibility of a fundamental, irreducible nature, which forces us to acknowledge that such trajectory in space does not exist, and since it does not exists, we must also abandon the idea that a microscopic entity would be always present in the 3-dimensional space.

In the words of the previous metaphor, a microscopic entity is essentially a diver, not a swimmer, and if you look for a diver you will find him almost always on the trampoline, or in the air, and not in the water. On the other hand, the human body, which is a macroscopic entity, is a genuine spatial entity, that is, a swimmer, not a diver, who cannot disappear from space as if by magic, only to reappear in another region of the same; certainly not in normal conditions, and according to the known laws of physics.

Le me add a further terminological clarification. In the scientific literature the term \emph{non-spatiality} is much less used than that of \emph{non-locality}. However, both terms express the same idea. In fact, all that is stably present in our three-dimensional space is necessarily local, that is, locally present in it, in \emph{actual} and not in \emph{potential} terms (also an extended object, like a cloud, is a local object, as it possesses local actual properties). Therefore, what is not present in a local sense is in fact not present at all (which doesn't mean it doesn't exist), and consequently the concepts of non-locality and non-spatiality are intimately related.

Now, as I tried to illustrate with the metaphor of the pool, reality is layered, and one of these layers is that in which the microscopic entities live: it is a layer that could be called \emph{prespatial} (and which in a sense is also \emph{pretemporal}). In adherence to this prespatial layer (represented in the metaphor as the layer of air above the pool), lies our ordinary spatial layer (the water in the pool), inside which macroscopic entities usually evolve, like the objects of our daily lives, and our human bodies. 

Of course, the pool metaphor is only to be understood as a very crude allegorical simplification. The non-spatial or prespatial layer is a non-ordinary reality whose dimensionality is much higher than the three dimensions of our ordinary space, or the four dimensions of spacetime, and in general it could even be considered to be infinite-dimensional (as infinite-dimensional is in general the Hilbert state-space of a quantum entity). This cannot be represented in the too simple pool metaphor, in which the dimensionality of the air region above the pool, and of the water region inside the pool, is the same. Also, the region of contact between the spatial and prespatial layers is much more articulate and intricate than what the metaphor suggests, and certainly the non-spatial (or pre-spatial) entities cannot be represented as simple corpuscular entities.

\section{Wave-packet spreading}

The problem that we need first to consider is the evolution of the \emph{probability of presence} (in space) of a microscopic entity, such as a single atom, when it evolves freely, i.e., when no external forces or other entities (microscopic or macroscopic) interact with it (apart the measuring system). The term ``probability of presence'' should be understood in the sense of the probability with which the microscopic entity in question lends itself to the creation of a spatial localization, in a given region of space $R$, at a given time $t$, through its interaction with a measuring apparatus~\cite{Aerts1999,Sassoli2012}.
In quantum theory, this probability is given by the squared modulus $|\psi_t({\bf x})|^2$ of the wave function (or wave packet) $\psi_t({\bf x})$ (describing the state of the entity in question, at time $t$), integrated over the spatial region $R$, that is: 
\begin{equation}
{\cal P}_t(R)=\int_R d{\bf x}\, |\psi_t ({\bf x})|^2.
\end{equation}
What I am now going to do is to estimate the width of such wave packet in the simplest case of a hydrogen atom, which is the first and simplest element of the famous Mendeleev's periodic table.

To determine the wave function of a hydrogen atom, it is useful to express the problem in the so-called variables of the \emph{center of mass} and \emph{relative movement}. In doing so, I will neglect for simplicity the description of the spins of the electron and proton. Without going into the details of this procedure, which can be found in any textbook of quantum mechanics, we can observe that due to this change of variables it is possible to transform the problem of two interacting bodies (electron + proton) into an effective, simpler problem, of two bodies that evolve independently of each other, and whose equations can therefore be solved separately.

The first body corresponds to the evolution of the center of mass of the system, and is equivalent to the evolution of a free entity (an entity evolving in the absence of any interaction) of total mass $M=M_e+M_p$, where $M_e$ and $M_p$ are the masses of the electron and proton, respectively. The second body corresponds instead to the evolution of a entity of (reduced) mass $\mu = M_e M_p/M$, which moves in the presence of a Coulombian central force field.

The solutions of the Schr\"odinger equation associated with the first problem are the so-called plane waves, which cover a continuum of possible energies, from zero to infinity (we speak in this case of a continuous spectrum). The solutions of the problem with the central force field are instead associated with discrete energy values, given by the well-known formula: $E_n= -E_I/n^2$, where $n=1,2,\dots$, and $E_I\approx 13.6\, eV\approx 22 \times 10^{-19}$~J is the ionization energy of the hydrogen atom. One speaks in this case of a discrete spectrum, corresponding to the known (emission and absorption) spectral lines, observed experimentally.

Now, as regards the possibility of acquiring different positions in space, what really matters is the movement of the center of mass of the hydrogen atom, which, as previously mentioned, evolves according to free evolution. What we are interested in is to calculate the spatial spreading of the wave packet associated with the center of mass variable, since such spreading will provide us a good estimate of the probability of observing the hydrogen atom at a certain distance from the place where it was initially observed, say at time $t = 0$~s.

The spatial spreading of the wave packet at time $t$ can be estimated by calculating the so-called \emph{standard deviation} $\Delta Q_t$ of the position observable ${\bf Q}$ (associated with the center of mass), which by definition is given by the square root: $\Delta Q_t =\sqrt{\langle {\bf Q}^2\rangle_t- \langle {\bf Q}\rangle_t^2}$, where $\langle\cdots\rangle_t$ denotes the quantum average relative to the state of the center of mass entity. Using \emph{Ehrenfest theorem}, with which one can calculate the average values of quantum observables, one can show (following a little long but not difficult calculation) that by judiciously choosing the origin of the time axis the spreading $\Delta Q_t$ of the center of mass wave packet, at time $t$, is given by:
\begin{equation}
\Delta Q_t = \sqrt{{\Delta P_0^2\over M^2} \, t^2 + \Delta Q_0^2},
\label{eq1}
\end{equation}
where $\Delta Q_0$ is the spatial spreading at time $t=0$, and $\Delta P_0$ the spreading with respect to the momentum observable ${\bf P}$, at time $t=0$.

For the initial width $\Delta Q_0$ of the packet, we can choose the typical value of the \emph{Bohr radius} (which in the semiclassical model of the Danish physicist corresponds to the radius of the innermost electron), i.e., about $5.3 \times 10^{-11}$~m ($0.53$~angstroms). For the value of $\Delta P_0$, we can instead consider a dispersion which is compatible with the energy of the ground state of the hydrogen atom, i.e., such that ${\Delta P_0^2\over 2M}\approx E_I$. Considering that the total mass is: $M \approx M_p \approx 1.67 \times 10^{-27}$~kg, we have $\Delta P_0 \approx 8.6 \times 10^{-23}$~Js/m, which is compatible with Heisenberg's principle, as is clear that: $\Delta Q_0\Delta P_0 \approx 45.6 \times 10^{-34}~{\rm Js}\approx 43\, \hbar > \hbar / 2$ ($\hbar \approx 1.05 \times 10^{-34}$~Js).

Inserting the above values in (\ref{eq1}), and observing that the second term in the square root is negligible compared to the first, we thus obtain $\Delta Q_t \approx 5.1 \times 10^4\, t$~m/s, that is: 
\begin{equation}
t\approx 0.2\times 10^{−4}\,\, \Delta Q_t ~{\rm s/m}. 
\label{eq1bis}
\end{equation}
This last expression tells us the time we roughly need to wait for the center of mass of the wave packet of the hydrogen atom to reach the spatial spreading $\Delta Q_t$. Let us consider some specific values. To obtain a spreading of $5$~km, that is, of $5 \times 10^3$~m, we have to wait about $10^{-1}$~s, i.e., a tenth of a second. In 1~s, instead, the packet will have reached a width of about $50$~km, while in $10$~s its approximate width will be of $500$~km, and so forth. In other words, the \emph{effective speed} with which the radial dimension of the center of mass wave packet grows, is approximately $50$~km/s, that is $180,000$~km/h, which is a speed of all respect, and corresponds, in our previous metaphor, to the maximum speed of the diver (from our ordinary spatial perspective this is however only a \emph{potential} speed, and certainly not an \emph{actual} speed, as it is not associated with a body moving through our ordinary space).

\section{Disassembling the body}

Summarizing, for a hydrogen atom we have determined the approximate behavior of that part of the wave function which describes the potential spatial localization of its center of mass. In doing so, we have ignored for simplicity the relative motion between the proton of the nucleus and the orbital electron, as well as their spins. More precisely, we have calculated how the width of the center of mass wave function varies over time in (configuration) space, due to the so-called phenomenon of the \emph{spreading of the wave function}, which can be understood as being a consequence of Heisenberg's uncertainty principle.

What is important to understand is that the domain in which the wave function is sensibly different from zero corresponds to the spatial region within which the atom in question has a chance of being detected. So, if the hydrogen atom, at time $t= 0$, was localized in a sphere whose radius is approximately equal to the Bohr radius, i.e., $r_0 = 5.3 \times 10^{-11}$~m (that is, its probability of presence in that sphere, at time $t=0$~s, is equal to $1$), what we have determined is that after for example $1$~s, that localization radius will have approximately grown to about $50~{\rm km} = 5 \times 10^4$~m.

What we are interested in is to estimate the probability with which we can detect the atom not in any location of this macro-sphere of $50$~km of radius, but in a predetermined sub-region of it. Indeed, if we later want to extrapolate our reasoning to an entire macroscopic structure, it is necessary that every atom forming the structure will re-locate in a very specific place in relation to all the other atoms of the structure, so as to reconstitute it in every detail. So, let us suppose that this sub-region corresponds to a micro-sphere whose radius is equal to the Bohr radius $r_0$. 

To estimate the above probability I will make an additional simplifying assumption. The wave function being not a constant function, the probability of presence will vary according to the location of the micro-sphere within the macro-sphere. However, since we are only interested in estimating a rough order of magnitude, we can assume that the wave function is a step-function, only taking two values: a constant non-zero value inside the macro-sphere, and a zero value outside of it.

With this simplification, we have everything we need to complete our estimation. For this, we have to remember that to calculate a probability of presence we have to integrate the squared module of the wave function over the region of interest. Considering the step-function hypothesis, this means that the probability that we seek will be proportional to the relative volume of the micro-sphere compared to the volume of the macro-sphere.

More exactly, given that the volume of a sphere is proportional to its radius to the cube, we obtain for the probability ${\cal P}$ that the hydrogen atom in question will be detected, after $1$~s, in a predetermined micro-sphere of radius $r_0$, within the macro-sphere of radius $r= 50$~km, the following order of magnitude:
\begin{equation}
{\cal P}\approx {r_0^3\over r^3}\approx \left({5.3\times 10^{-11} \,{\rm m}\over 5\times 10^4 \,{\rm m}}\right)^3\approx 10^{-45}.
\label{eq2}
\end{equation}
This is undoubtedly a very small number, with 45 zeros after the decimal point! And of course, we can easily do the same calculation for larger macro-spheres, i.e., waiting more time than just a second. For example, if we wait $10$~s, the radius of the macro-sphere will increase by a further factor of 10 (from $50$~km to $500$~km), and consequently the estimated value of the probability ${\cal P}$ will decrease from $10^{-45}$ to $10^{-48}$, i.e., by a factor of a thousand, and so on.

Now that we have obtained an estimate of the probability of quantum self-teleportation of a hydrogen atom from an initial micro-sphere to a given final micro-sphere, we must consider the case of an entire macroscopic body, like that of a human being of planet Earth, which we can assume to have a mass of $100$~kg. Here of course we have to face the already mentioned problem that a macroscopic body has the property of spatiality, and therefore cannot be conveniently described by a wave function (more will be said about this in the next section). But suppose that for some reasons, unknown to us, all the interatomic bonds suddenly cease to exist, so that in an instant all the atoms that form the human body in question become separate and independent from each other, bringing them back to the prespatial layer of our physical reality. 

On the basis of this sci-fi hypothesis, each individual atom of the body structure can be conveniently described by the laws of quantum mechanics, and we can apply the previous calculation to each one of them. In fact, this is not really true, as is clear that all these atoms will constantly be bombarded by the countless entities present in the environment, in particular the thermal photons, so that we also need to assume that the sci-fi process of disassembly of the human body is able to induce a perfect isolation of the different atomic constituents from all the other entities (micro and macro) present in the environment. 

To keep the discussion as simple as possible, we further assume that the body structure is constituted solely by hydrogen atoms, and since the mass of a hydrogen atom is about $1.67\times 10^{-27}$~kg, a human of $100$~kg, if constituted only by hydrogen atoms, would contain approximately a number $N$ of them given by:
\begin{equation}
N\approx {100\, {\rm kg}\over 1.67 \times 10^{-27}\, {\rm kg}}\approx ≈ 4.2 \times 10^{28}.
\label{eq3}
\end{equation}

Each of these atoms will have to individually re-locate in a specific region of space, to reconstitute the entire body structure, with no errors. Therefore, the (estimated) probability ${\cal P}_{\rm body}$ of self-teleportation of the overall body structure will be given by the product of the self-teleportation probabilities ${\cal P}$ of every single atom contained in that structure. If the body would be formed only by two atoms, that is, $N= 2$, the probability would be ${\cal P}_{\rm body} \approx {\cal P}^2\approx 10^{-45\times 2} = 10^{-90}$. With three atoms, the probability would become: ${\cal P}_{\rm body} \approx {\cal P}^3\approx 10^{-45\times 3} = 10^{-135}$. Therefore, with $N= 10^{28}$ atoms, we obtain:
\begin{equation}
{\cal P}_{\rm body} \approx {\cal P}^{10^{28}}\approx 10^{-4.5\times 10^{29}}. 
\label{eq4}
\end{equation}

Let us reflect for a moment on the amazing infinitesimality of this number. To write it in decimal, non-scientific notation, we must use more than $10^{29}$ zeros, i.e., more than one hundred billion billion billion zeros! If we write with a printer on paper ten zeros per second, to write the entire number will take us more than $10^{28}$ seconds, that is, more than $10^{21}$ years, which is about a hundred thousand billion times the assumed age of the known universe (according to today cosmological theories)! In other terms, the value we have obtained for ${\cal P}_{\rm body}$, although not strictly equal to zero, is nevertheless so small that we have no point of comparison to be able to understand it. Yet, we have probably overestimated it.

In fact, we have assumed that for a reason unknown to us all the atoms of the human body will suddenly disassemble and become non-spatial entities, so allowing their individual wave packets to spread. But we have also neglected the problem of the relative motion between the different atomic constituents, equating the individual atoms to free elementary-like particles. Furthermore, we have neglected the spin variables of the different atomic constituents. Also, we have assumed that the environment in which the different atomic components evolve, once disassembled, corresponds to an effective absolute vacuum, otherwise the associated wave-packets cannot be considered to evolve freely, and that each atom is able to evolve without interacting with all the others, before regaining a specific spatial location. In addition to that, we have hypothesized that when the various atoms reappear in the relative positions they occupied before being disassembled, the entire macroscopic structure will be able to reconstitute, without any particular inconvenience. Taking into account all these assumptions would of course further reduce the value of ${\cal P}_{\rm body}$, by a factor which is very difficult, if not impossible, to evaluate.

But that's not all. There is another ``sci-fi miracle,'' which is implicit in our reasoning, perhaps even more amazing than that of the disassembly of the initial structure (which is possible to relax; see the next section). This second miracle has to do with the different hydrogen atoms being simultaneously ``drawn'' back into space. Let me explain. An elementary entity, such as a proton, an electron, or an entire hydrogen atom, spends most of its time in a non-spatial (non-local) condition, unless it is incorporated into a macroscopic structure. Now, although there is no consensus on this among physicists, many agree that a microscopic entity is unable to acquire a precise spatial localization spontaneously, as this can only be done by interacting with a macroscopic material structure, like for instance that forming a measurement apparatus. 

Experimental physicists are undoubtedly able to build detection apparatuses allowing microscopic entities to temporarily acquire a spatial localization, in specific places, and even though these apparatuses could in principle localize in space a certain number of microscopic entities at a time, as far as I know a device which can localize an entire macroscopic structure doesn't exist, and perhaps is not even conceivable. The possibility remains, of course, that the process of spatial localization could occur even in the absence of macroscopic structures playing the role of detection devices, as is suggested in some interpretations of quantum theory, like the so-called objective collapse theories~\cite{Ghirardi2011}, the transactional interpretation of quantum physics~\cite{Ruth2013}, and others, the discussion of which, however, would go beyond the scope of the present article. 

Among the factors that we have not taken into account in the estimation of ${\cal P}_{\rm body}$, there are of course also those that could slightly increase the value of the probability. For example, we have implicitly assumed that all the atomic components have to re-localize at exactly the same instant. However, nothing prevents us from admitting a small time-delay in the localization process of the individual atoms, which, if sufficiently small, may not affect the correct re-assembly of the entire body macrostructure. But it is unlikely that considerations of this kind would be able to significantly change the infinitesimality of ${\cal P}_{\rm body}$.

\section{Cooling down the body}

At this point some readers may rightly object that we don't really needed our ``disassembling sci-fi hypothesis,'' as what generally makes a macroscopic object like our human body behave classically, i.e., spatially and locally, is just the fact that it is immersed in a thermal environment, i.e., that it is constantly subjected to the random collisions of countless microscopic entities, in particular photons, and that the overall effect of these innumerable interactions is that of producing its continuous ``collapse into space,'' which would be essentially the reason why it would behave differently from a ``pure'' quantum entity, like an electron. To use once more our metaphor, this bombardment is what would force the body to remain inside the water of the pool, preventing its owner from becoming a diver. 

So, one could object that, to allow the body to quit the ``spatial pool,'' and temporarily become a non-spatial entity, it would be sufficient to shield it from the external thermal environment, so that there would be no need for having it first disassembled into smaller atomic fragments and then recombined, which is the operation that apparently produced the inconceivable infinitesimality of the self-teleportation probability, as each of the $10^{28}$ fragments needed to relocalize in a predetermined place, within a sphere of $50$~km of radius. 

This is a pertinent objection that I'm now going to explore. This objection, by the way, could appear to be in contradiction with what I have just stated above, at the end of the last section, regarding the lack of an apparatus that could objectify (spatialize) a whole macroscopic structure. If our standard terrestrial environment is able to keep a macroscopic body into space, then wouldn't be that same environment the measuring apparatus that is able to achieve the required goal of producing the collapse -- the objectification -- of an entire macroscopic object? 

If this is true, then it would be sufficient to isolate an ordinary object to obtain its automatic de-localization (i.e., its de-spatialization). However, we cannot  expect this to work, as the object is also in contact with another environment: its own \emph{internal} one. If the body is sufficiently large, as is certainly the case of a human body (but also of a speck of dust, and of much smaller entities), then the mutual interactions of its constituents can also have an influence in determining its overall classical (spatial) versus quantum (non-spatial) behavior. 

The reason for this is easy to explain. In the case of the hydrogen atom, we were able to separate the wave function relative to the center of mass from that associated with the relative motion. In this way, the center of mass was described by a free evolving wave function. With a macroscopic body, we may want to do the same, i.e., to separate the wave function describing the center of mass from the contribution coming from the different movements of all its constituents, relative to that center and to each others. Here we can consider the ensemble of these constituents as an entity playing the role of a measuring apparatus with respect to the ``center of mass entity,'' so that the latter would be constantly subjected to a measurement process, thus producing its classical behavior. 

Therefore, to describe the center of mass by means of a free evolving wave function, the evolution of the body's center of mass needs to decouple from that of its internal degrees of freedom, and this can reasonably be done only if the body is cooled down to extremely low temperatures. How low? Well, we can say, remaining here necessarily vague, low enough to avoid any exchange of energy between the center of mass degree of freedom and the degrees of freedom associated with the internal relative movements~\cite{Sun2001}.

In the previous section we have assumed that by some sci-fi action the body was all of a sudden disassembled (and each constituent isolated from one another, and the environment). This was an assumption of simplicity, as in this way we were able to use the well-known factorization of the wave-function of a two-body system, which of course is much harder to obtain in general for a macroscopic body. We can however replace the ``disassembling sci-fi hypothesis'' with the requirement that not only the body in question will have to evolve in a perfect vacuum (no thermal bombardment), but also that it will be cooled down instantaneously to temperatures almost equal to the absolute zero. 

In other terms, we now replace the ``disassembling sci-fi hypothesis'' by a ``freezing sci-fi hypothesis.'' The advantage is that cooling down a body seems an operation less impossible to achieve than disassembling it into its atomic fragments, without destroying them, but also, and more importantly for our idealized discussion, the entire structure of the body will be preserved in this way, which hopefully will increase the value of the self-teleportation probability. Of course, also in this case we will have to assume that the body is additionally isolated from the environmental thermal bombardment. 

So, let us assume, as we did before, that the mass of the body is $10^2$~kg. By assumption, since the internal and external environments have now been made totally silent, and cannot anymore play the role of generalized detector instruments with respect to the wave function of the body's center of mass, we can consider that the latter is described by a free evolving wave packet. To further simplify the discussion, we assume that at time $t=0$, the wave packet is approximately Gaussian (this is a reasonable assumption, considering that the probability density of any non-Gaussian wave packet becomes approximately Gaussian as it spreads~\cite{Mita2007}), which means that the inequality in Heisenberg's uncertainty principle is approximately an equality: $\Delta Q_0\Delta P_0 \approx {\hbar\over 2}$. 

As we did with the hydrogen atom, we take the standard deviation of the center of mass position observable $\Delta Q_0$ to be equal to the Bohr radius. Therefore:
\begin{equation}
\Delta P_0\approx {\hbar\over 2 \Delta Q_0}\approx {1.05 \times 10^{-34}\, {\rm J}\, {\rm s} \over 2 \times 5.3 \times 10^{-11}\, {\rm m}}\approx 10^{-24}\, {\rm kg}\, {\rm m}\, {\rm s}^{-1}.
\label{eq5}
\end{equation}
Inserting this value into (\ref{eq1}), we find that the spreading $\Delta Q_t$ of the center of mass wave packet at time $t$ is given by:
\begin{equation}
\Delta Q_t \approx \sqrt{(10^{-52} \,{\rm m}^2 \,{\rm s}^{-2})\, t^2 + 3\times 10^{-21}\,{\rm m}^2}.
\label{eq6}
\end{equation}

Now, the original question of the sci-fi writer was to have the body disappearing from one place and reappearing tens of kilometers away. Considering, as we did before, a distance of $50$~km, we have to set $\Delta Q_t = 50~{\rm km} = 5\times 10^{4}~{\rm m}$ in the above equation. If we do so and solve for $t$, we find the value $t\approx 5\times 10^{30}$~s. Considering that 1 year corresponds to $3.154\times 10^7$~s, we obtain that the center of mass wave packet of the macroscopic body will reach a width of $50$~km after approx. $1.6\times 10^{23}$~years, which is approximately ten million billion times the assumed age of the known universe! 

Here we see an important difference between the wave packet spreading of a hydrogen atom, who was extremely fast, and the wave packet spreading of a macroscopic body, which is inconceivably slow. But to answer the question of the sci-fi writer, we certainly cannot wait so long, because he explicitly asked the self-teleportation to happen in a matter of seconds. Also, even in case we would accept to wait for so long, we may have a problem with the ``expiration date'' of our universe! And anyhow, without having to freeze the body and the environment, using any classical means of transport through space (the ``swimming modality'') would be in this case much more effective to travel the distance of $50$~km. 

On the other hand, we can also say that, although the width of the wave packet almost doesn't increase as time passes by, even if we perfectly confine the position of center of mass in a given small region of space, at time $t=0$~s, a fraction of a second after its wave function will have acquired an infinite tail. The value of this tail will be infinitesimally small, but nevertheless different from zero. So, let us estimate this value, at a distance of $50$~km, after exactly one second of free evolution. 

The body's center of mass wave packet can be written as the product of three identical Gaussian factors: $|\psi_t({\bf x})|^2= |\psi_t(x_1)|^2 |\psi_t(x_2)|^2|\psi_t(x_3)|^2$, where: 
\begin{equation}
|\psi_t(x_i)|^2=\sqrt{2\over\pi a^2}{1\over \sqrt{1 + {4\hbar^2 t^2 \over M^2a^4}}}\exp\left[-{{2a^2\left(x_i-{\hbar k_i\over M}t \right)^2}\over a^4 + {4\hbar^2 t^2 \over M^2}}\right],\quad i=1,2,3.
\label{gaussian}
\end{equation}
To evaluate $|\psi_t({\bf x})|^2$, at time $t=1$~s, we can set $k_1=k_2=k_3=0$ (the body is at rest at time $t=0$~s), $x_2=x_3=0$~m, $x_1\approx 5\times 10^4$~m, $M=10^2$~kg, and $a = 2r_0 \approx 10^{-10}$~m. This implies that $|\psi_1(x_2)|^2=|\psi_1(x_3)|^2\approx 10^{10}$, and $|\psi_1(x)|^2\approx 10^{10}\times e^{-5\times 10^{29}}$. Multiplying the probability density $|\psi_t({\bf x})|^2$ by the volume of the micro-sphere of Bohr radius in which we want the center of mass to relocate, at time $t=1$~s, we thus obtain that: 
\begin{equation}
{\cal P}_{\rm body} \approx e^{-5\times 10^{29}}\approx 10^{-2.2\times 10^{29}}. 
\label{eq4-bis}
\end{equation}

Comparing (\ref{eq4-bis}) with (\ref{eq4}), we observe that have obtained a self-teleportation probability of the same order of magnitude. This means that, quite surprisingly, even if we avoid the sci-fi procedure of disassembling the macroscopic body, which as we have seen was responsible for the inconceivable infinitesimality of the obtained probability, and replace it with a procedure of total internal freezing, thus preserving the structural integrity of the body, a similar inconceivably infinitesimal self-teleportation probability is obtained, this time because of the extreme slowness of the spreading of the macroscopic wave function and the extreme infinitesimality of its long-distance tails.

\section{Superselection rules}

Considering my above pessimistic analysis, I'll leave it to the science fiction writer the task of finding a convincing sci-fi solution, not violating too many physical laws at the same time, allowing the hero of his story to teleport himself and accomplish his mission, whatever it is. As for me, let me offer a final thought. 

According to quantum theory, and the phenomenon of the spreading of the wave packet, a hydrogen atom, if left to evolve freely, will quickly acquire a truly gigantic size, apparently in contradiction with what is usually observed. Also, when considering the spectrum of energies of a hydrogen atom, in addition to the discrete energy values, associated with the relative electron-proton movement, we also have to consider the continuous energy values associated with the translational degrees of freedom of the center of mass. The spectrum of the total energy of the atom is thus given by the sum of these two energy spectra. But the sum of a discrete spectrum and a continuous spectrum produces a continuous spectrum, apparently in contradiction with the spectral lines experimentally observed.

In short, without further precautions, the application of the Schr\"odinger equation to the problem of the hydrogen atom does not allow to obtain results in agreement with the experimental observation, that is, in agreement with the fact that atoms do not usually possess macroscopic sizes, nor spectra of a continuous nature. To solve this problem, one can make use of the notion of \emph{superselection rules}~\cite{Streater1964}. Rules of this kind restrict the physically realizable states, and when associated with a given observable, they prevent considering states that would be a superposition of states associated with different values of this observable, as these superpositions would be in disagreement with the experimental data. In other terms, the existence of superselection rules indicates that the structure of the state space is not strictly Hilbertian (as linearity would not apply for all states), but more general. 

An example of superselection rule is that associated with the observable determining whether the infamous Schr\"odinger's cat is alive or dead. If $\psi_A$ is the wavefunction describing the alive cat, and $\psi_D$ the wave function describing the dead cat, then, as far as we know, the wave function $\psi = \psi_A + \psi_D$ obtained by superposing these two wave functions does not describe a physically realizable state. This means that there is a superselection rule on the ``life observable'' of the cat, which forbids the superposition of wave functions characterized by different values of this observable.

In the case of the hydrogen atom, if we want to obtain values for its energy spectrum in agreement with the experimental data, it is necessary to consider the position and momentum of its center of mass as variables of a classical kind, associated with superselection rules, and same thing if we want to correctly describe its observed non-macroscopic size. Of course, the reasons for this  inhibition of quantum  superpositions and the associated classical behavior of certain observables can be multiple, and will generally depend on the specificities of the environment in which the entity in question is immersed. So, determining what are the classical observables and what the quantum ones, in a given context, is a problem not necessarily easy to solve, and there is no unique recipe for this: in some contexts, certain observables will behave classically, while in other contexts they will behave quantum mechanically, and  still in others their behavior will be semiclassical, or semiquantum, that is, in between these two regimes. 

But then, if the center of mass of the hydrogen atom is the expression of a superselection rule, goodbye self-teleportation! On the other hand, if we consider it as quantum observable, goodbye agreement with many experimental data. But as I said, to determine the classical or quantum nature of an observable it is necessary to take into account the specificities of the experimental context. When an atom is incorporated into a macroscopic material structure, or undergoes continuous interactions with countless microscopic entities and force fields present in the environment, it usually undergoes a process of de-synchronization of its wave function, able to transform certain quantum observables into classical ones. That's why, in the beginning of this article, I have argued that, strictly speaking, quantum self-teleportation would be impossible. More precisely, it is impossible (${\cal P}_{\rm body}=0$) if we consider the standard environment in which we humans  evolve, which makes our bodies, and the objects with which we interact in our everyday life, classical entities.

What I'm here suggesting is that the classical or quantum nature of a physical entity is not an intrinsic feature of the same, but a contextual one: in some contexts certain entities will behave as quantum entities (when subjected to certain observational processes), and in other contexts they will behave, instead, classically. These considerations open to an important reflection, which can be summarized in the following question: 
\\

\noindent \emph{Is the physical reality fundamentally quantum?}
\\

\noindent The majority of physicists seem to believe so, that is, to think that quantum theory would be more fundamental than classical theory, and that a classical behavior would always emerge from a quantum substrate, when certain circumstances are met. However, a different view is also possible. For instance, one can consider that our physical reality is neither classical nor quantum, but genuinely hybrid, that is, a complex combination of these two aspects. 

In other words, the physical entities forming our reality would generally be quantum-like, i.e., they would be entities potentially manifesting both aspects, the classical and the quantum aspects, depending on their state and the nature of the experimental questions we address to them. According to this view, supported by some very general (operational) approaches to the foundations of physical theories, especially that of the so-called Geneva-Brussels School of Quantum Mechanics (nowadays mainly active in Belgium, at the Center Leo Apostel, led by the Belgian physicist Diederik Aerts), the classical regime and the quantum regime would correspond to very specific limit cases of more general situations~\cite{Aerts1998,Aerts1999,Sassoli2013,Aerts2014d}.

More precisely, the classical regime would be associated with experimental situations where all fluctuations can be controlled, so that all observational processes are predictable in advance. On the other hand, the quantum regime would  be associated with experimental contexts in which the fluctuations are maximal (and uniform), so producing a situation of maximum lack of knowledge. In between these two regimes, intermediate, hybrid regimes can also exist, neither purely classical nor strictly quantum, that physicists have just begun to investigate and that seem to provide a more general and realistic model for the description of the countless physical entities in interaction with their multiple environments.


\begin{thebibliography}{}
\bibitem{Bennet1993} C. H. Bennett et al., ``Teleporting an Unknown Quantum State via Dual Classical and Einstein-Podolsky-Rosen Channels,'' Phys. Rev. Lett., 70, 1895--1899 (1993).
\bibitem{Wootters1982} W. Wootters and W. Zurek, ``A Single Quantum Cannot be Cloned," Nature 299: 802--803 (1982).
\bibitem{Dieks1982} D. Dieks, ``Communication by EPR devices,'' Physics Letters A 92(6): 271--272 (1982). 
\bibitem{Ma2012} X. Ma, T. Herbst, T. Scheidl, D. Wang, S. Kropatschek, W. Naylor, B. Wittmann, A. Mech, J. Kofler, E. Anisimova, V. Makarov, T. Jennewein, R. Ursin and A. Zeilinger, ``Quantum teleportation over 143 kilometres using active feed-forward,” Nature 489, 269--273 (2012).
\bibitem{Aerts1998} D. Aerts, ``The entity and modern physics: the creation discovery view of reality,'' In: \emph{Interpreting Bodies: Classical and Quantum Objects in Modern Physics}, edited by Elena Castellani (pp. 223--257). Princeton Unversity Press, Princeton (1998).
\bibitem{Aerts1999} D. Aerts, ``The Stuff the World is Made of: Physics and Reality,'' In: \emph{The White Book of `Einstein Meets Magritte'}, edited by Diederik Aerts, Jan Broekaert and Ernest Mathijs (pp. 129--183). Kluwer Academic Publishers, Dordrecht (1999).
\bibitem{Sassoli2011} M. Sassoli de Bianchi, ``Ephemeral Properties and the Illusion of Microscopic Particles,'' Foundations of Science 16, 393--409 (2011).
\bibitem{Sassoli2012} M. Sassoli de Bianchi, ``From Permanence to Total Availability: A Quantum Conceptual Upgrade,'' Foundations of Science 17, 223--244 (2012).
\bibitem{Sassoli2013} M. Sassoli de Bianchi, ``The $\delta$-Quantum Machine, the $k$-Model, and the Non-ordinary Spatiality of Quantum Entities,'' Foundations of Science 18, 11--41 (2013).
\bibitem{AertsSassoli2014} D. Aerts and M. Sassoli de Bianchi, ``The extended Bloch representation of quantum mechanics and the hidden-measurement solution to the measurement problem,'' Annals of Physics 351,  975--1025 (2014).
\bibitem{Sassoli2015} M. Sassoli de Bianchi, ``God May Not Play Dice, But Human Observers Surely Do,''  Foundations of Science 20, 77--105 (2015).
\bibitem{Neumann1932} J. Von Neumann, ``Grundlehren,'' Math. Wiss. XXXVIII (1932). 
\bibitem{Bell1966} J. S. Bell, ``On the Problem of Hidden Variables in Quantum Mechanics,'' Rev. Mod. Phys. 38, 447--452 (1966). 
\bibitem{Gleason1957} A. M. Gleason, ``Measures on the closed subspaces of a Hilbert space,'' J. Math. Mech. 6, 885--893 (1957). 
\bibitem{Jauch1963} J. M. Jauch and C. Piron, ``Can hidden variables be excluded in quantum mechanics?,'' Helv. Phys. Acta 36, 827--837 (1963).
\bibitem{Kochen1967} S. Kochen and E. P. Specker, ``The problem of hidden variables in quantum mechanics,'' J. Math. Mech. 17, 59--87 (1967). 
\bibitem{Gudder1970} S. P. Gudder, ``On Hidden-Variable Theories,'' J. Math. Phys 11, 431--436 (1970).
\bibitem{Ghirardi2011} G. C. Ghirardi, ``Collapse Theories,'' The Stanford Encyclopedia of Philosophy (Winter 2011 Edition), Edward N. Zalta (ed.), URL = <http://plato.stanford.edu/archives/win2011/entries/qm-collapse/>.
\bibitem{Ruth2013} R. E. Kastner, \emph{The Transactional Interpretation of Quantum Mechanics}, Cambridge University Press, Cambridge (2013).
\bibitem{Sun2001} C. P. Sun, X. F. Liu, D. L. Zhou and S. X. Yu, ``Localization of a macroscopic object induced by the factorization
of internal adiabatic motion,'' Eur. Phys. J. D 17, 85--92 (2001).
\bibitem{Streater1964} R. F. Streater and A. S. Wightman, \emph{PCT, spin and statistics, and all that}, W. A. Benjamin, Inc., New York (1964). 
\bibitem{Mita2007} K. Mita, ``Dispersion of non-Gaussian free particle wave packets,'' Am. J. Phys. 75, 950--952 (2007).
\bibitem{Aerts2014d} D. Aerts, ``Quantum Theory and Human Perception of the Macro-World,'' Front. Psychol. 5, article 554 (2014).



\end{thebibliography}
\end{document}